\documentclass[conference]{IEEEtran}

\usepackage{cite}
\usepackage{amsmath,amsfonts}
\usepackage{amssymb}
\usepackage[T1]{fontenc}
\usepackage{algorithm}
\usepackage{algpseudocode}
\usepackage{graphicx}
\usepackage{textcomp}
\usepackage{xcolor}
\usepackage{braket}
\usepackage{tabularx}
\usepackage{booktabs} 
\newcommand{\locoswitch}{LoCoSwitch}
\newcommand{\locoswitchfull}{\underline{Lo}gical-level, \underline{Co}de-combining \underline{Switch}}

\def\BibTeX{{\rm B\kern-.05em{\sc i\kern-.025em b}\kern-.08em
    T\kern-.1667em\lower.7ex\hbox{E}\kern-.125emX}}

\begin{document}

% Ensure letter paper
\pdfpagewidth=8.5in
\pdfpageheight=11in

\pagenumbering{arabic}

%%%%%%%%%%%%%%%%%%%%%%%%%%%%%%%%%%%%

\title{A quantum store-and-forward switch achieving order-of-magnitude throughput gains from combining logical ebits on atom-cavity arrays}
\author{
\parbox{\textwidth}{\centering
Willers Yang\textsuperscript{1,*},
Nayana P. Tiwari\textsuperscript{2,3,*},
Tina Oberoi\textsuperscript{1},
Dahlia Ghoshal\textsuperscript{2,3},
Justus C. Brüggenjürgen\textsuperscript{3},
Alexander R. Korsch\textsuperscript{3,4},
Carl Meeuw\textsuperscript{3},
Hannes Bernien\textsuperscript{3,4},
Frederic T. Chong\textsuperscript{1}\\[0.6em]
\small
\textsuperscript{1}\textit{Department of Computer Science, University of Chicago,} Chicago, IL, USA\\
\textsuperscript{2}\textit{Pritzker School of Molecular Engineering, University of Chicago,} Chicago, IL, USA\\
\textsuperscript{3}\textit{Institute for Experimental Physics, University of Innsbruck,} Innsbruck, Austria\\
\textsuperscript{4}\textit{Institute for Quantum Optics and Quantum Information of the Austrian Academy of Sciences,} Innsbruck, Austria\\[0.3em]
\textsuperscript{*}These authors contributed equally to this work.
}
}

\maketitle
%%%%%%%%%%%%%%%%%%%%%%%%%%%%%%%%%%%%

\begin{abstract}
Fault-tolerant application scale quantum (FASQ) computation requires a quantum error correcting code (QECC) to reach target fidelity, yet quantum networks are built to route physical entangled pairs (ebits), with each logical qubit costing a couple to over hundreds of physical ebits depending on the protocol. 
We present a quantum store-and-forward network switch that pools and routes logical ebits rather than physical ones: it combines the logical ebits of connected nodes into high-rate QECCs chosen per link and traffic class via code-switching, performs error-correction and swaps logical ebits in batch. 
The design of this switch spans three levels: at the physical level, we give the design and modeling of a hardware on a cavity-coupled neutral-atom array that achieves storage, computation and entanglement generation on a single device; at the logical level, an explicit code-switching protocol based on ultra-high-rate codes and surgery; and at the network level, a logical entanglement routing protocol that can be distributed, along with a logical network simulator and performance analysis on real and synthetic traces. 
On the Abilene backbone with 30 users per city and 1\% Bell-pair infidelity, our protocol sustains 25.8$\times$ the capacity of the baseline that routes physical ebits at the FASQ target of $10^{-12}$ and 56.2$\times$ at the early fault-tolerant target of $10^{-6}$ with 3\% infidelity, 25.7$\times$ and 27.2$\times$ that of a repeater that can distill every link, and 2.0$\times$ to 58.3$\times$ across 12 settings over three topologies.
Overall, we achieve for entanglement routing what store-and-forward switching did for classical data packets: by pooling, re-encoding, and routing logical ebits in batch, it lets many users share a scarce link more efficiently, so a physical link with the same ebit generation rate can carry an order of magnitude more logical traffic.
\end{abstract}
\section{Introduction}\label{sec:intro}

\begin{figure}[t]
\centering
\includegraphics[width=\linewidth]{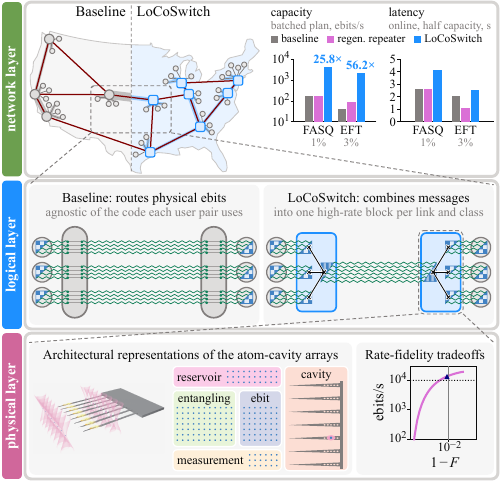}
\caption{\locoswitch{} at three levels, read from the bottom up. \textit{Physical:} an atom--cavity node that stores, computes and generates ebits in one system, at $10^{4}$ ebits per second at 99\% fidelity. \textit{Logical:} a code-switching protocol that combines the logical ebits of users sharing a next hop into a more ebit-efficient high-rate code. \textit{Network:} a routing algorithm that chooses a path with a QEC protocol per link, leading to 25.8$\times$ to 56.2$\times$ higher logical throughput over baseline on Abilene.
}\label{fig:hero}
\end{figure}

Quantum networks are central to both scaling quantum computation and realizing quantum communication. For computation, fault-tolerant application-scale quantum (FASQ) computers \cite{preskill2025megaquop} capable of factoring~\cite{shor1994algorithms} and quantum simulation~\cite{feynman1982simulating} are expected to scale as distributed systems over many modules, machines, and clusters~\cite{smith2022chiplet,XueCovey2026,kim2024million}, connected via a quantum network. For communication, the quantum network distributes entangled pairs (\emph{ebits}) through repeaters \cite{briegel1998repeaters,Azuma2023QuantumInternet} and enables protocols with provable guarantees, such as unconditionally secure key distribution~\cite{ShorPreskill2000BB84} and data verification~\cite{anschuetz2022supercheq}. It is therefore critical for a quantum network to deliver ebits at both high enough rates and fidelity to enable quantum advantage.

Existing quantum hardware~\cite{Bluvstein2024LogicalArrays} and links \cite{hartung2026quantumnetworkingspeedquantum, Krutyanskiy2024MultimodeFiber}, however, operate at physical error rates orders of magnitude higher, often trading physical resources such as physical qubits and ebits for improved fidelity. Indeed, while the demonstrated error rates of ebits across a noisy repeater are at a few percent~\cite{main2025dqc}, applications need logical error rates (LER) from about $10^{-6}$ for early fault-tolerant (EFT) demonstrations \cite{Katabarwa2024EFTQC} to less than $10^{-12}$ per program qubit inside FASQ applications such as factoring~\cite{Beverland2022Assessing,gidney2021factor}. Quantum error correction (QEC) and distillation protocols can close the fidelity gap by encoding logical information redundantly over many noisy physical resources~\cite{Calderbank_1996,bennett1996mixed}, with an overhead that may depend on hardware capabilities, logical targets, and the number of qubits it encodes in a batch \cite{xu2024constant,bravyi2024high}. 

\begin{figure}[t]
\centering
\includegraphics[width=\columnwidth]{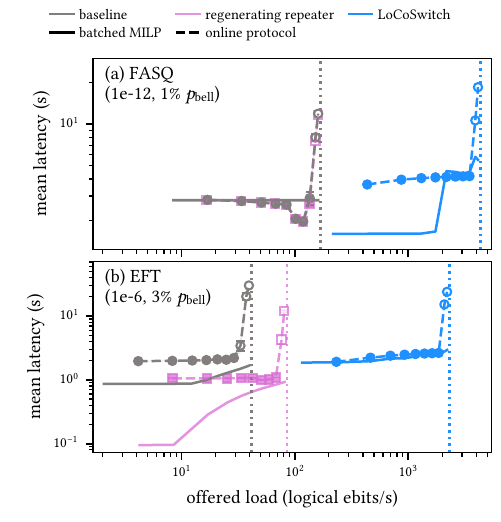}
\caption{Load-latency frontier on Abilene with 30 users per city, at (a). FASQ target LER $10^{-12}$ with $1\%$ $p_{bell}$ and (b). EFT target LER $10^{-12}$ with $1\%$ $p_{bell}$; towards lower right is better. 
}
\label{fig:hero-frontier}
\end{figure}
We use the \emph{logical throughput} of a network, or the number of logical ebits it delivers per second at target logical error rates, as the key performance metric. 
It can be improved in two ways: by lowering the cost of a logical ebit via more efficient error correction protocols\cite{muralidharan2018oneway,nishio2025multiplexed,pattison2025constant,bonillaataides2025constant} or by delivering physical ebits more efficiently across a network \cite{Sinclair2025Fault-tolerantArrays,shi2020concurrent,pant2019routing,pirker2018modular,wallnofer2019multipartite,huang2025peer}, and most existing work considers these two directions separately. 
In this \emph{compartmentalized} view, which we use as a baseline, user nodes choose their error-correction protocols end-to-end, and the physical ebits they request are routed by a network that is agnostic of the encoding. Although it keeps the network simple, it overlooks critical optimizations. First, it requires all pairs of end nodes to coordinate a protocol directly, whereas a growing network would prefer simpler and more distributed controls. Secondly, end-to-end physical ebits can accumulate costs in fidelity or rate that grow unfavorably with the number of hops, which means future generation repeaters such as \emph{regenerating repeaters} will already need to perform QEC in addition to the QEC that happens at end nodes \cite{Azuma2023QuantumInternet}.
Finally, the network does not benefit from more efficient encoding when there exists one, for example, when the same destination logical qubits can be batched into high-rate codes with better overhead. These costs are largest where some links in the network are scarcer or more congested than others, which we observe both in a distributed quantum computer where communication between clusters are much slower than between the modules in the same cluster \cite{yoder2025tour,wegmann2026chipmunq,ang2024arquin}, and also in large-scale entanglement distribution, where rate and fidelity of ebit generation both fall with distance~\cite{pirandola2017fundamental,Azuma2023QuantumInternet}, making inter-city links more scarce than metropolitan (metro) links.  

An analogous problem is dealt with in classical networks by a switch that receives traffic from many links, buffers it, and forwards it by destination and class~\cite{enrightjerger2017onchip,kleinrock1975queueing,kleinrock1978principles,rfc2475}. Message combining switch on shared-memory multiprocessors, which merges the requests of many processors for the same memory location into one message as they cross the network~\cite{gottlieb1983ultracomputer,dickey1992combining}, uses a similar idea to reach throughput gains.  
Three important gaps remain in building a quantum network switch: at the physical level, hardware that can store and compute on logical qubits while generating entanglement with its neighbors at a high rate; at the logical level, a way to code-switch quickly and fault-tolerantly on the same hardware; and at the network level, a protocol that distributes logical entanglement and determines the optimal encoding used, ideally one that is distributed and scalable. We present \locoswitch, a \locoswitchfull\, that addresses these gaps and routes logical ebits rather than physical ones, giving quantum networks fault-tolerant, distributed entanglement routing that makes efficient use of its scarce bandwidth. (See Figure~\ref{fig:hero} for the design of \locoswitch\ in the three layers)
In designing \locoswitch, we make four contributions:

\begin{enumerate}
\item \textbf{New hardware and logical designs (Section~\ref{sec:methods}).} We give a node design and logical protocols that together implement a message-combining quantum network switch, and give a detailed model using experimental data, projected device parameters, and circuit-level simulations.
\item \textbf{Logical entanglement routing algorithms. (Section~\ref{sec:framework}).} We formulate entanglement routing over error correction protocols as a maximum concurrent flow and give two solutions: as an exact mixed-integer linear program (MILP), the batched algorithm, which approximates the optimal logical throughput gain, and as an online, decentralized protocol that runs on a small request header without a solver.
\item \textbf{A quantum network simulator.} We release \emph{locosim}, an event-driven simulator of logical routing whose routing and pooling rules are plug-ins of about 30 lines, validated against closed forms within 3\%.
\item \textbf{Improving logical throughput of quantum networks. (Section~\ref{sec:evaluation}, Figure~\ref{fig:hero-frontier}).} On the Abilene backbone with 30 users per city and 1\% $p_{\rm bell}$, the batched algorithm gives \locoswitch{} 25.8$\times$ to 56.2$\times$ the baseline's logical throughput depending on the target LER, or 25.7$\times$ and 27.2$\times$ that of a regenerating repeater that distills every link. 
The gain comes from pooling on scarce links, largest where the intercity links are slowest and where many users share a switch, and it holds across 12 topologies considered (2.1$\times$ to 29.6$\times$ at FASQ and 2.0$\times$ to 58.3$\times$ at EFT). 
\end{enumerate}

Together, these results show that routing logical ebits through switches that choose the encoding per link and class can raise the logical throughput of a rate-bottlenecked quantum network without raising the rate at which its links generate physical ebits. 
The rest of the paper is organized as follows: Section~\ref {sec:background} reviews quantum networks in three layers, with related work; Section~\ref{sec:methods} describes the architecture of \locoswitch{} in detail, with the routing algorithms in Section~\ref{sec:framework}; Section~\ref{sec:evaluation} introduces \emph{locosim}, an event-driven quantum network simulator which we used to evaluate \locoswitch, and Section~\ref{sec:conclusion} concludes with limits and open problems.

\section{Background}\label{sec:background}
This section reviews what a FASQ network requires at each of the three levels drawn in Figure~\ref{fig:hero}. At the network level, users request logical entanglement at target error rates and the network routes it; at the logical level, error-correcting codes protect the quantum information that this entanglement moves; and at the physical level, links generate entanglement that nodes store and compute on. Most existing designs leave the choice of these codes to the end users or choose one family of protocol for the network to improve fidelity, whereas \locoswitch{} chooses, for each logical request, a plan of QEC protocols chosen at each link based on physical links below as well as on the network traffic above. We proceed down the stack from the network to the hardware levels, then discuss the classical networks that inform our design and conclude with a review of related research.
\subsection{Quantum Networks}\label{sec:bg-qnet}

The basic unit of quantum information is a qubit, and the unit of entanglement is an \emph{ebit}, or a maximally \underline{e}ntangled pair of qu\underline{bits} that can be distributed to distant users over a quantum network \cite{kimble2008}.  Once two nodes share an ebit, a quantum teleportation operation can transfer the state of a qubit from one node to the other by consuming the ebit, two classical bits~\cite{bennett1993teleporting,hermans2022teleportation}, and some local operations. The primary task of a quantum network is to efficiently supply ebits between the users at required fidelity.

A link in the quantum network can supply noisy \emph{physical ebits}, most often by entangling local qubits with flying photonic qubits, which are then interfered or absorbed to extend this local--flying entanglement to entanglement between distant nodes~\cite{zukowski1993event,Pan1998ExperimentalInteracted}. In the DLCZ approach~\cite{Duan_2001}, a single detector click after a beam splitter heralds the entanglement, giving a high success rate but requiring the optical phase of the fiber link to be stabilized. Two-photon schemes~\cite{Barrett_2005} instead interfere one photon from each node and herald on two detector clicks, removing the phase-stability requirement at the cost of a success probability $\propto \eta^2/2$ for a linear-optics Bell-state measurement (BSM). Alternatively, one node can emit the photon and the other can directly absorb it with a time-reversed pulse sequence~\cite{Cirac1997StateTransfer}, heralded by a non-destructive check of the receiving qubit, which avoids the factor of 1/2 from the BSM. Ebits between nodes that are not directly connected can be created by \emph{entanglement swapping}, where an intermediate node connected to both users consumes its two pairs to entangle the two neighbors, with an error roughly equal to the sum of the errors of the links it joins. Because these errors accumulate along a route, a network may also distill its ebits, trading several noisy ebits for fewer, higher-fidelity Bell pairs~\cite{bennett1996purification,DEJMPSprotocol,zhu2024squeezing}. Since fiber loss grows exponentially with distance, repeater nodes are placed along long links to join and distill ebits over shorter segments.

\subsection{Quantum Error Correction}\label{sec:bg-qec}
An end user in the network may request ebits at logical error rates (LER) much lower than what the network can supply physically, in which case quantum error correction (QEC) protocols based on a quantum code is used to trade many copies of noisy physical ebits with fewer, higher-fidelity ones. We can group a user's requests by their error target into three \emph{traffic classes}, from those used for fault-tolerant application scale quantum application (FASQ class) with target LER of $10^{-12}$ \cite{Beverland2022Assessing,preskill2025megaquop}, to early fault-tolerant demonstrations (EFT class) with target LER of $10^{-6}$ \cite{Katabarwa2024EFTQC}, to LERs compatible with near-term intermediate noisy hardware (NISQ class) at target LER of $10^{-3}$. Requests of every class are served by \emph{logical ebits}, entangled Bell pairs between logical qubits of two code blocks, which are prepared from physical ebits at a cost that depends on the code. When logical ebits are prepared transversally~\cite{stack2026transversal}, a code block consumes one physical ebit per physical qubit, shared among the logical qubits that the block encodes. A logical ebit thus can cost anywhere between over a hundred physical ebits in a surface-code patch, which holds one logical qubit, to only two in a filled block of a rate-1/2 code. 

A QEC protocol is described by its code parameters $[[n,k,d]]$, where $n/k$, the number of physical qubits used ($n$) over the number of logical qubits encoded ($k$) determines the overhead of the protocol, and the distance $d$ determines the minimum number of errors it can tolerate before an undetected logical error occurs. 
In the traditional approach based on surface codes that protect a single logical qubit, this overhead can grow quadratically with the number of correctable errors, leading to resource overheads on the order of 100-1000 physical qubits per logical qubit to reach FASQ fidelity targets \cite{Beverland2022Assessing,gidney2021factor}. Recent progress in quantum low-density parity check (qLDPC) codes have found families of protocols encoding a batch of logical qubits in the same block with a constant overhead \cite{xu2024constant, bravyi2024high} that can typically improve as we increase the number of logical qubits encoded together, with the most efficient family costing about only $2$ physical qubits per logical qubit on average over a batch of hundreds of logical qubits \cite{kasai2026breaking,yang2026designercodesgalacompact,lee2026designprinciplesultrahighratequantum}.

\subsection{Experimental Realization}
A logical entanglement routing and switch node must combine reconfigurable, scalable, high-fidelity computing with high-rate, high-fidelity entanglement generation. Neutral atom arrays are currently the most scalable quantum computing platform, with defect-free, long-lived arrays of thousands of qubits~\cite{Manetsch2025}, increasingly high single- and two-qubit gate fidelities~\cite{evered2026highfidelityentanglinggatesnonlocal}, and the logical operations required for early fault-tolerant computation~\cite{Bluvstein2024LogicalArrays}. On the networking side, generating ebits at high rate over telecom fiber requires a strong light--matter interface~\cite{kimble2008}; recent demonstrations across atoms and ions are summarized in Ref.~\cite{menssen2026strategicplanneutralatom}. The strength of this interface is characterized by the cooperativity,
\begin{equation}
    C = \frac{4g^2}{\kappa\gamma} \propto \frac{\lambda^3 Q}{V},
    \label{coop_eqn}
\end{equation}
where $g$ is the atom--cavity coupling rate, $\gamma$ is the atomic decay rate, and $\kappa = \kappa_{wg} + \kappa_{sc}$ is the total cavity decay rate, the sum of decay into the waveguide and scattering into free space. Equivalently, $C$ scales with the quality factor $Q$, a measure of how long light is trapped in the cavity, and inversely with the mode volume $V$. High cooperativity, $C \gg 1$, enables both high-fidelity atom--photon entanglement and efficient photon collection.

Remote matter--matter entanglement mediated by cavity-generated photons has been demonstrated across several platforms~\cite{ritter2012elementary, knaut2024entanglement, Krutyanskiy2023EntanglementMeters, Chiarella2025ANetworking}. Macroscopic mirror cavities in atom~\cite{Chiarella2025ANetworking, Shaw2025AInterfacing, desantis2026realizationcavitycoupledrydbergarray} and ion systems~\cite{fabitha2026trappedionbasedquantumrepeaternode, Canteri2025GenerationNode} have shown time-multiplexed entanglement generation, but their large mode volumes limit $C$ and thus the per-attempt success probability. Fiber cavities currently reach the highest demonstrated cooperativity, $C = 100$~\cite{Grinkemeyer2024Error-DetectedCavity}, and both atom and ion platforms have shown temporal multiplexing by cycling atoms through the cavity mode~\cite{roschinski2026programmablecavityqedfiberintegrated, Canteri2024ANode, Hu2025Site-SelectiveRegister}. However, with a single cavity per node, the number of entanglement attempts within each herald window is limited, which caps the entanglement rate. Often, quantum frequency conversion to telecom reduces this rate further.

Nanophotonic crystal cavities offer a path beyond this limit. They achieve the smallest mode volumes, yielding high $g$, and can be engineered for strong waveguide coupling $\kappa_{wg}$~\cite{Menon2024AnImaging, Dordevic2021EntanglementTweezers, ozawa2026strongcouplingreconfigurable171yb}. With the ability to fabricate hundreds of cavities on a single millimeter-scale chip, the scalability combined with ideal cavity parameters makes the platform well-suited for spatially multiplexed, high-rate entanglement generation. Loading, imaging, and rearranging atom arrays next to such a chip have been demonstrated~\cite{Menon2024AnImaging}, but simultaneous operation of many cavities at state-of-the-art cooperativity has yet to be realized. A recent architecture proposal combines fault-tolerant neutral-atom computing with cavity-based interconnects and predicts MHz rates between co-located modules~\cite{Sinclair2025Fault-tolerantArrays}. We extend this direction to a full quantum network: we propose physical-layer designs for user, switch, and repeater nodes built on spatially multiplexed nanophotonic cavities, specify their ebit-generation protocol, and calculate ebit rates over local and 500\,km links at $p_\mathrm{bell}$ target infidelities.

\subsection{Classical Networks and Related Work}\label{sec:related}
The scenario of many local devices connected sharing a link is analogous to classical networks with network switches. In \emph{store-and-forward packet switching}, a switch buffers the packets of many flows whose traffic arrives in bursts and forwards them over one shared link, allowing the link to be sized for the combined traffic of the flows rather than for the sum of their peaks~\cite{baran1964distributed,kleinrock1975queueing}. Dragonfly interconnects likewise route the traffic of many cheap local channels over a few costly global ones~\cite{kim2008dragonfly}. 
Combining switches go one step further and merge the messages themselves: in the NYU Ultracomputer, a switch that holds two requests for the same memory location forwards one combined request and splits the reply on its way back, so that many processors can access one location at once instead of queuing at its memory module~\cite{gottlieb1983ultracomputer}, and pairwise combining switches implement this in the processor-memory network of its prototype~\cite{dickey1992combining}. 
Such networks are typically evaluated by their throughput, the latency of their slowest packets, and loss; by the utilization of their expensive links; and by the cost per delivered bit~\cite{kleinrock1978principles}. Reliability, in contrast, is left to the end users, who recover a lost packet by resending it from the sender's copy~\cite{saltzer1984endtoend}. 

Most quantum network proposals treat the network as a service that delivers physical ebits at a requested fidelity. Network stacks take requests that give the target fidelity~\cite{dahlberg2019link,kozlowski2020designing}, and routing treats all physical ebits as interchangeable~\cite{chakraborty2020multicommodity,shi2020concurrent,pant2019routing,li2024minerva}. Switches~\cite{vardoyan2021stochastic,zhang2025switchqnet,sharma2026performance}, data-center networks and interconnects~\cite{shapourian2025qdc,sakuma2024qfly}, network processors~\cite{li2025qnpu}, and compilers of distributed programs~\cite{wu2023qucomm,wu2022autocomm,xiang2025adaptdqc} likewise switch, buffer, or protect physical ebits. Fault-tolerant end users then protect their own data end to end on top of this service~\cite{pattison2025constant,stack2026transversal,kim2024million}. Our baseline follows this practice, which keeps the network simple but misses critical optimization opportunities for routing physical ebits agnostic of the encoding that consumes them.

There exist quantum networking protocols that use distillation or error correction for routing ebits at improved fidelity. For example, encoded repeaters protect pairs in transit using a choice of code specified by a hardware regime~\cite{jiang2009repeater,fowler2010surface,muralidharan2016optimal}, or using surface codes of different distances~\cite{stack2026transversal, zhang2026spatiotemporal}; and ReFINE schedules distillation and coding together inside the network, trading rate against fidelity~\cite{alavisamani2026refine}.  Routing has likewise been extended to encoded qubits assuming a uniform family of codes~\cite{hu2024surfacenet,fan2026scope}. These protocols differ from our approach in that they target fidelity of the delivered ebit instead of throughput and do not pool or combine across multiple messages. Finally, while code-switching between low- to high-rate codes is studied in compilation and memory hierarchy studies~\cite{wu2023changeable,stein2025hetec,viszlai2026qsieve}, they have not been studied in the context of entanglement routing or network design. Unlike these designs reviewed above, \locoswitch\ treats the network neither as a service that delivers physical ebits nor as a chain of repeaters that protect each pair in transit. Its switches instead route logical ebits, choosing a code for each link and traffic class and combining the logical ebits of many users into shared blocks. They thus play the role of the classical switch described above, but they also correct errors in what they forward and re-encode it for the next link. These switches do not, however, raise the rate at which the links generate physical ebits, and their gain comes at the cost of memory and latency, since the pools of logical ebits that they keep occupy atoms and require additional time to perform code-switching and decoding (Section~\ref{sec:evaluation}). To measure this gain, we compare \locoswitch\ with the baseline and with stronger designs that distill or encode inside the network.

\label{sec:logical}
\begin{figure}
    \centering
    \includegraphics[width=\linewidth]{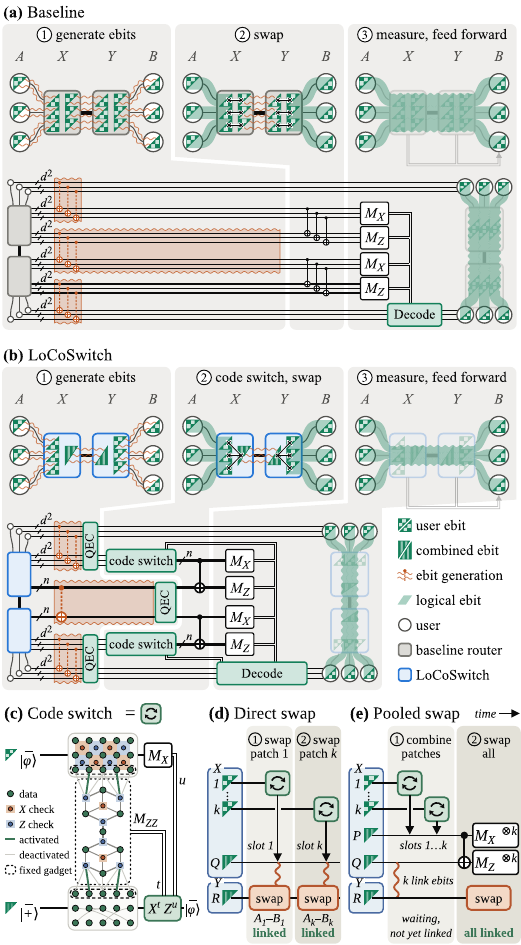}
    
    \caption{The logical implementation of \locoswitch. (See Section~\ref{sec:logical-layer} for a detailed description.)} 
    \label{fig:logical}
\end{figure}

\section{LoCoSwitch Architecture}\label{sec:methods}
The design of LoCoSwitch spans the physical, logical, and network layers, where the protocol at each level hands a few parameters to the next (Figure~\ref{fig:hero}). The physical level protocol defines each link's rate, in physical ebits per second, its infidelity $p_{\rm bell}$, and each switch's atom and cavity budgets, and its physical error model. The logical level protocol gives a menu of codes, described by its overhead in terms of physical ebits per logical ebit, its logical error rates, the number of logical qubits it encodes, and its \emph{storage limit}, or the longest the block can protect encoded logical qubits against a total error budget, and the costs of code-switching in both space-time volume and logical errors. Finally, the network-level protocol defines the logical routing problem with requests grouped into classes, solved optimally as a mixed-integer linear program (MILP) or heuristically as a distributed streaming protocol. 

\subsection{The Physical Layer}
For a realizable implementation of \locoswitch, we consider network nodes based on a neutral-atom array coupled to an array of high-cooperativity nanophotonic devices. In this architecture, we trap the atom array with tightly focused optical tweezers, a platform that enables quantum computation with initialization, single-qubit gates, and two-qubit gates, plus networking capabilities.

Each nanophotonic device consists of a silicon nitride photonic crystal cavity tuned to atomic resonance, and a waveguide that shapes the emitted photon into a fiber-compatible mode. We fabricate the devices to suspend from a millimeter-scale silicon chip like an array of diving boards~\cite{Menon2024AnImaging} (Figure~\ref{fig:hero} physical layer), and we trap atoms a few hundred nanometers above each device surface to evanescently couple to the cavity mode. In addition to spatial multiplexing across the cavity array, we interleave several atoms in the gaps between devices, so a single batch of atoms delivered to the cavity zone can make repeated entanglement attempts like a conveyor belt. The cavity array emits photons that are then coupled into a waveguide-to-fiber array converter outside the vacuum chamber, enabling multiplexing, routing, and in situ rerouting of collected photons to various local users or to distant nodes.

Using this strong light--matter interface in the over-coupled regime ($\kappa_{wg} > \kappa_{sc}, g$), we can generate high-fidelity atom--photon entanglement. Here, we consider cesium, which has an excited-state transition at 1359.2\,nm in the telecom O-band. For lower loss in fiber networks, the same entanglement scheme and atom array capabilities can be adapted to rubidium~\cite{ShankarGirijavallabhanMenon2025CombiningNetworks}, which has an excited-state transition within the lowest-loss telecom C-band. However, more detailed consideration of the atom--photon entanglement scheme would be required due to additional hyperfine energy levels in the excited states.

As depicted in the physical layer of Figure~\ref{fig:hero}, the node includes a cavity networking zone, an entangling zone, a measurement zone, and a reservoir zone, consistent with current state-of-the-art atom-array experiments~\cite{Chiu_2025, Manetsch2025, Bluvstein2024LogicalArrays}. Atoms from the reservoir are shuttled on and off the cavities in parallel conveyor-belt moves to attempt ebit generation. Heralded ebits are then moved to the entangling zone, located approximately 200\,$\mu$m from the devices to mitigate the effects of stray charges on high-fidelity Rydberg gates~\cite{Ocola2024ControlDevice}, while a second conveyor brings new atoms to the cavity zone. In the entangling zone, each batch of link ebits is purified and, at repeaters, swapped with ebits connected to neighboring links, extending entanglement across longer distances. A single node that combines quantum computation with high-rate ebit generation at high fidelity is suitable to support the logical and networking protocols in this work.
\subsection{The Logical Layer}\label{sec:logical-layer}
In this section, we describe how a physical network node given above is used to implement the logical primitives required. We consider the case where users are connected to each other through a network of repeaters and switches (Figure~\ref{fig:logical}a-b). The logical protocol is described in Figure~\ref{fig:logical}, and it requires the following logical capabilities:

\noindent\textbf{\textcircled{\scriptsize 1} Preparing logical ebits.} Logical ebits can be prepared transversally~\cite{stack2026transversal, bonillaataides2025constant,pattison2025constant}. To do so, we first prepare two blocks of $[[n,k,d]]$ code in $|+\rangle_L$ and $|0\rangle_L$ initial states, at the two ends of a link. Then, by generating and consuming $n$ physical ebits, we can implement $n$ physical CNOTs across the link (the dashed orange gates in stage~1 of Figure~\ref{fig:logical}a and~b, labeled by the number of ebits they consume -- $d^2$ in an example baseline using surface codes or $n$ for the QEC protocol \locoswitch\, picks), which results in a logical CNOT between the $k$ logical ebits, entangling them across the link. Both ends then extract syndromes and compare them (the green QEC boxes in Figure~\ref{fig:logical}b). What follows depends on the code; an error-detecting code, such as the $[[17,9,4]]$ code from \cite{pattison2025constant}, keeps the block only if the syndromes agree and discards it otherwise, while an error-correcting code, such as a high-rate LP code\cite{bonillaataides2025constant}, decodes the difference and keeps the block. The two can be concatenated, with a detection level on top of decoded blocks. A block costs $n$ physical ebits, or $n/k$ ebits per logical ebit: $d^2=121$ for a $d=11$ surface-code patch and 2 for a filled GALA-672 block. Discarded blocks add to this: a $[[17,9,4]]$ detection level keeps 0.846 of its injected ebits and 0.807 of its blocks, so it costs $17/(9\cdot 0.846\cdot 0.807)=2.77$ physical ebits per logical ebit rather than $17/9$. Note that L1 is the only operation that consumes physical ebits.

\noindent\textbf{\textcircled{\scriptsize 2} Logical Pauli measurements and Code-switching.} Users hold halves of logical ebits in their own code, here surface-code patches, while the shared link carries a high-rate block. A switch therefore has to move a logical qubit from an $[[n_1,k_1,d_1]]$ code into an $[[n_2,k_2,d_2]]$ code, and back on the far side. We call this operation code-switching (the ``code switch'' boxes in stage~2 of Figure~\ref{fig:logical}b). We build this from two logical measurements: a joint $\bar Z\bar Z$ between a surface-code patch and one logical qubit of a high-rate block, performed by high-rate surgery~\cite{maskara2026efficient}, and transversal $\bar X$ or $\bar Z$ readouts. To switch codes, the switch measures $\bar Z\bar Z$ between the user-side patch and a logical qubit of the switch block, and then reads out the patch in $\bar X$. This moves the logical state from the patch into the block, up to a Pauli correction fixed by the two outcomes. Figure~\ref{fig:logical}c shows this circuit at the logical level: the $\bar Z\bar Z$ parity $t=m_a\oplus m_b$ is read out through a helper pair $(a,b)$ held inside the switch, the $\bar X$ readout gives $u$, and the state appears in the block up to $X^tZ^u$. The figure draws this correction explicitly, but in the protocol no outcome is corrected in place: all outcomes are decoded together and fed forward as a single Pauli correction at the receiving user (``Decode'' in stage~3 of Figure~\ref{fig:logical}b). \emph{Cost.} A code switch is local to the switch and consumes no physical ebits. It costs time and adds a logical error given in Figure~\ref{fig:betas}.

\noindent\textbf{\textcircled{\scriptsize 3} Logical entanglement swap.} A switch holds one half of an ebit shared with a user and one half of an ebit shared with the next switch. A logical Bell measurement on the two halves, a joint $\bar Z\bar Z$ followed by $\bar X$ readouts of both, joins the two ebits into one longer ebit. When the two halves sit in different codes, this measurement also performs the code switch. \locoswitch\ does the swap in two ways, trading off memory and latency overhead. In the space-efficient \emph{direct swap} (Figure~\ref{fig:logical}d), each user's patch is measured against its own logical qubit of the link's block $Q$ as soon as it arrives, so no user waits for another and only a few user ebit is stored at the switch each time. In the time efficient \emph{pooled swap} (Figure~\ref{fig:logical}e), users' halves are first loaded by L3 into a second, local block $P$ (step~1) via high rate surgey, and then a transversal CNOT onto the link's block followed by transversal readouts then swaps every logical qubit of the two blocks at once (step~2). In both, the link ebits connect $Q$ to its partner block $R$ at the far switch, which swaps each logical qubit of $R$ with the patch of its own user (``swap at $Y$'') and so completes the end-to-end ebit. The direct swap gives each user the shortest wait and suits light traffic. The pooled swap holds waiting users in a block rather than in patches, about $n/k$ physical qubits each instead of $d^2$, and touches the link's block in a single step, so it suits heavy traffic. Both variants take two joint surgeries, one at each switch, and $d^2+n/k+d^2$ physical ebits per end-to-end logical ebit, of which only $n/k$ falls on the scarce link; the pooled swap adds one block of memory at the switch. Section~\ref{sec:evaluation} returns to when combining users pays.

Once the end-to-end logical ebit exists, a user teleports its data over it. The sender applies a local transversal CNOT from its data patch onto its half of the ebit, both held in its own code, reads out the two patches, and sends the two outcomes classically. The receiving user combines them with the outcomes fed forward from the switches and applies a single Pauli correction. Only classical bits pass between the two users at this stage, and the data never passes through a switch.

\subsection{The Network Layer}\label{sec:framework}

We can define the network as a graph $G=(V,E)$ with user and switch nodes, and links $e$ that generates physical ebits at rate $R_e$ with infidelity $p_{{\rm bell},e}$, modeled after the physical layer design, and each end of a link has $D_e$ decoders and switch $v$ an optional atom budget $M_v$. The requests of one class between the user pairs behind one pair of switches form a \emph{flow} $u$ with demand $\lambda_u$ logical ebits per second.

The baseline chooses its protocol end to end among a surface-code patch, a BB-144 block~\cite{stack2026transversal}, or a high-rate qLDPC codes \cite{bonillaataides2025constant}, using raw ebits or over-distilled ebits~\cite{pattison2025constant}. The number of ebits required to implement the distillation and the code is then sent to the network, which routes physical ebits as a Multi-commodity max flow \cite{shahrokhi1990maximum,chakraborty2020multicommodity}. As a second baseline, we consider a \emph{regenerating repeater}, where every link may distill its pairs before the swap, so the end users get physical ebits that accumulate the errors of distilled rather than raw ebits.

\locoswitch{} chooses a \emph{plan} per request and pools at every link. The plan tells the network how a request is met: the path it takes, what each link along it turns its physical ebits into, and the QEC protocol used. Because its path and QEC protocols are set, a plan has a fixed error, summed over its code-switches, entanglement swaps, and storage time, and it is \emph{feasible} if that sum meets its class target. We build a flow's plans on its fewest-hop paths and up to two more, giving each hop the cheapest QEC protocol whose error fits its share of the budget. Online, the switch where a request enters picks its plan from a table and writes the plan's ID into the header, so every switch on the way knows which pool to claim from with no further decision.

The logical routing problem is then a maximum concurrent flow over plans rather than paths. If the traffic is known in advance, the problem can be solved centrally as a mixed-integer linear program (MILP). In practice, the network must serve requests as they arrive. As an approximation, we give the batched plan that waits for a window of time $W$, so the demand over the window is known and can be met optimally as the following linear program. Given $G$, the flows $u$ with their feasible plans $\mathcal L_u$ and a window $W$, choose plan flows $x_\ell\ge0$, a fraction $\Theta\ge0$, and for every pool $q$ integers $y_q,v_q\ge0$ and a rate $s_q\ge0$, to maximize $\Theta$ subject to
\begin{align}
&\textstyle\sum_{\ell\in\mathcal L_u}x_\ell=\Theta\lambda_u &&\forall u,\label{eq:dem}\\
&\textstyle W\sum_{\ell\ni q}x_\ell\le K_q\,(y_q+v_qW/\tau_q)+Ws_q &&\forall q,\label{eq:places}\\
&\textstyle P_e=\sum_{q\ni e}n_qB_q+W\sum_{\ell\ni e}c_{\ell,e}\,x_\ell\le R_eW &&\forall e,\label{eq:raw}\\
&\textstyle\sum_{q\ni e}\delta_qB_q+W\sum_{\ell\ni e}d_{\ell,e}\,x_\ell\le D_eW/t_{\rm dec} &&\forall e,\label{eq:dec}\\
&\textstyle\sum_{e\ni v}P_e/(R_eW)\le1 &&\forall v,\label{eq:cav}
\end{align}
where $\delta_q=1$ if pool $q$ makes blocks of a code that needs a decoder and $0$ otherwise. Constraint~\eqref{eq:dem} gives every flow the same fraction $\Theta$ of its demand, and \eqref{eq:places} covers the outputs that the plans using a pool take in a window. Constraint~\eqref{eq:raw} keeps the physical ebits a link spends, $P_e$, within those it generates; \eqref{eq:dec} keeps the blocks to decode within its $D_e$ decoders of $t_{\rm dec}$ per block, and \eqref{eq:cav} shares a switch's cavities among its links. A pool per user pair, the baseline's, and that of a check at the users, has no variables of its own: a plan pays there $c_{\ell,e}$ physical ebits and $d_{\ell,e}$ decodes per logical ebit, at the pool's expected fill under Poisson arrivals. 

The \emph{online protocol} determines the plan for a request at each switch from a static table and sends the header on. The table is built once from the long-run traffic matrix: for each flow, one plan per enumerated path, and the plan of least cost summed over its links fills the flow's demand. The online algorithm must also decide the window to wait for messages to accumulate, and trade the delay of a request against the number of requests that share a block: the longer its window, the more requests each block collects and the higher the encoding efficiency. How long to pool before code-switching is performed is determined using a streaming algorithm that prepares a block once the summed waiting of the queued requests reaches the cost of a block. This reaches a competitive factor of $2$; that is, the latency incurred is at most twice the optimum for a sequence of requests in hindsight.
\section{Evaluation}\label{sec:evaluation}
In this section, we evaluate the proposed physical, logical, and networking protocols of \locoswitch{} with experimental numerics and data, circuit-level simulations, and \emph{locosim}, an event-driven network simulator. First, we present an experiment proposal capable of achieving ebit generation rates and fidelities necessary to implement the logical layer. Then, we perform logical-level simulations to benchmark the logical error rate performance of message combining and logical entanglement swapping, before combining the resulting models into \emph{locosim} to test \locoswitch{} against the baseline and on real and synthetic traffic on 12 network topologies.

\subsection{Ebit Generation}

\begin{figure}[ht]
\includegraphics[width=0.95\columnwidth]{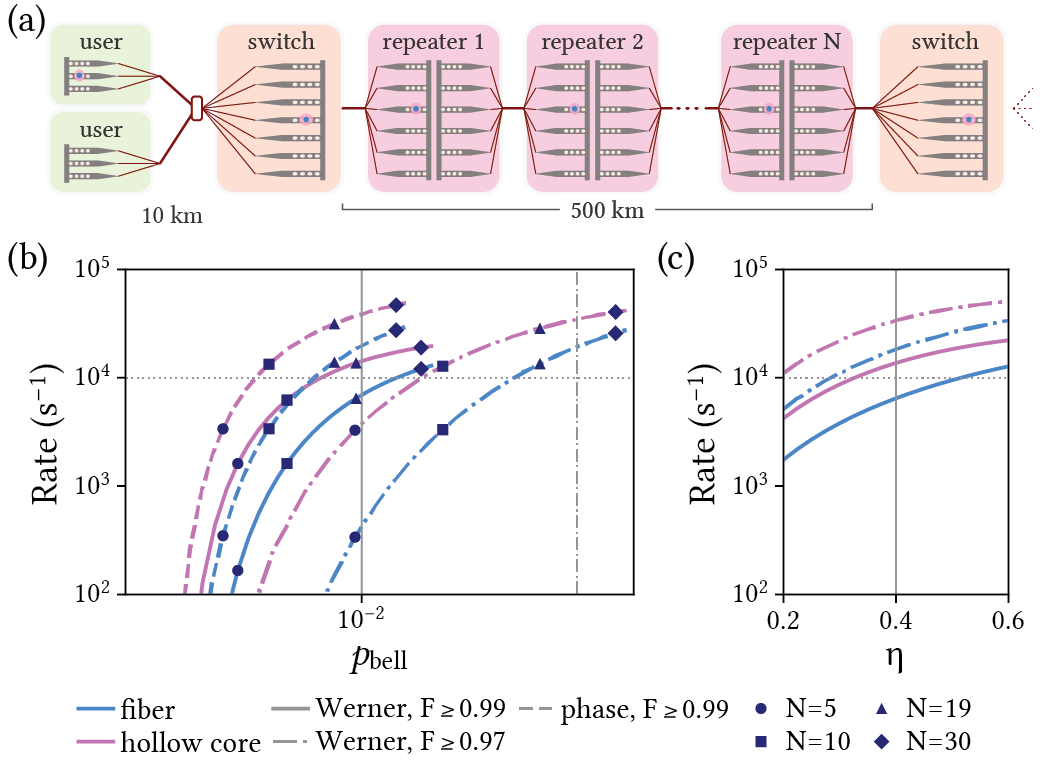} \caption{Entanglement generation in the proposed network. (a) Network schematic (b) Ebit rate versus $p_{bell}$ for targets $ 3\times10^{-2}$ and $ 10^{-2}$. (c) Ebit rate versus photon efficiency $\eta$}\label{fig:exp_numerics}
\end{figure}

In this network architecture, we generate entanglement local links, which connect users to \locoswitch{} over approximately 10\,km of fiber, and a 500\,km long-distance link between \locoswitch{} nodes in different metropolitan areas with $N$ repeater nodes interleaved, as depicted in Figure~\ref{fig:exp_numerics}a.  A link attempt succeeds with probability $p = \eta^2 T(d)$, where $T(d)$ is the fiber transmission over link length $d$ and $\eta$ is the photon collection efficiency.

Locoswitch and user nodes, which prioritize atom number, use $p_{2Q} = 10^{-3}$, consistent with the current state of the art (99.85\% raw, 99.94\% with atom-loss postselection~\cite{evered2026highfidelityentanglinggatesnonlocal}).  With only $p_{2Q} = 10^{-3}$ at every node, however, $p_{bell} \leq 10^{-2}$ is reachable only with at most two repeaters, and the resulting rate falls two to four orders of magnitude below the target. Thus, repeater nodes, which only perform purification and swapping on a comparatively small number of atoms, are assumed to reach $p_{2Q} = 10^{-4}$, for example by converting leakage into detectable erasures~\cite{Chow2024Circuit-basedAtoms}. To reach $p_{bell} \leq 10^{-2}$, we use two rounds of DEJMPS purification per link~\cite{DEJMPSprotocol}, accepting the lower rate due to consuming lower-fidelity ebits in the protocol. We assume White Rabbit synchronization on all fiber connections, providing sub-nanosecond timing between distant emission and absorption attempts~\cite{amiesking2025unrepeatedwhiterabbittime}. Additionally we consider both standard telecom fiber with approximately 0.2\,dB/km loss in the C-band, and hollow-core fiber, which has been demonstrated with less than 0.1\,dB/km loss~\cite{petrovich2025broadband, Boddeda_2025}. Since light travels at nearly $c$ in hollow-core fiber, this choice also reduces the classical bottleneck, which is the time each atom waits for a herald notification.

We take the infidelity of each heralded atom--photon emission or absorption to be $p_{atom-photon} = 10^{-2}$, based on previous proposals of time-bin protocols in similar systems~\cite{Menon2020NanophotonicInterface, Huie2021MultiplexedCavities}, and model the source of infidelity as general Werner noise or, optimistically, as pure phase error. We assume over-coupled cavities with cooperativity $C \approx 432$ coming from design parameters and a total photon collection efficiency per interface of $\eta = 0.4$, based on proposal estimates (Table~\ref{tab:assumptions}). Our model does not consider the following effects: memory decoherence, atom loss and heating during transport, scattering on idle atoms, and fiber connection losses.

\subsubsection{Long-distance link.}   Repeater and \locoswitch{} nodes are modeled with 150 devices with a 10\,$mu$m pitch with 5 atoms to attempt per cavity spaced at 2\,$mu$m; half the cavities are dedicated to each long-distance link neighbor(s) and the remaining at \locoswitch{} are for local users. As shown in Figure~\ref{fig:exp_numerics}b, there is a trade-off between rate and $p_{bell}$: as we increase the number of repeaters, we shorten each link and raise the rate, but each swap adds error. With the general Werner noise assumption, we reach the target $p_{bell} \leq 10^{-2}$ at the target rate of $10^{4}$\,ebits/s using hollow-core fiber with $N = 19$ repeaters ($1.4\times10^{4}$\,ebits/s, 25\,km links), while standard fiber reaches $6.5\times10^{3}$\,ebits/s. If we assume $p_{atom\text{-}photon}$ is due only to phase error, both fiber and hollow-core links meet the targets, with a minimum of 16 and 9 repeaters, respectively, due to only requiring a single round of link-to-link entanglement purification. For the relaxed target $p_{bell} \leq 3\times10^{-2}$, the final purification round between the \locoswitch{} nodes is no longer needed, and both channels exceed $10^{4}$\,ebits/s even with Werner noise: $1.8\times10^{4}$\,ebits/s over standard fiber and $3.4\times10^{4}$\,ebits/s over hollow-core fiber with $N = 23$ repeaters. Temporal multiplexing with five atoms per cavity increases the rate by $3.5\times$. Additionally, rather than fixing cavity-to-cavity chains across the link, pooling heralded ebits from all cavities increases the rate by a further $\sim$6$\times$.

\subsubsection{Local links.} In the local case, users have only 75 cavities with double the spacing, thus double the number of interleaved atoms, using two entanglement emission attempts per entanglement absorption at the \locoswitch{}. Dedicating 75 of its cavities to a single user over 10\,km of fiber, the \locoswitch{} generates ebits with $p_{bell} \leq 3\times10^{-2}$ at $4.7\times10^{5}$\,ebits/s over hollow-core fiber ($3.5\times10^{5}$\,ebits/s over standard fiber) without any purification, so only 17--22 cavities are needed to reach $10^{5}$\,ebits/s. For $p_{bell} \leq 10^{-2}$, two purification rounds are required (for Werner noise), and the same 75 cavities yield $8.3\times10^{4}$\,ebits/s over hollow-core fiber ($7.6\times10^{4}$\,ebits/s over standard fiber). Using all 150 of its cavities (e.g., two 75-cavity users), the \locoswitch{} reaches $0.9\times10^{6}$\,ebits/s at $p_{bell} \leq 3\times10^{-2}$ over hollow-core fiber, and $10^{6}$\,ebits/s with any one modest improvement: an additional shuttling of atoms, local links less than 5\,km, an $\eta = 0.45$, or faster conveyor transport. A future study could consider scheduling and allocating cavities to users and the long-distance link dynamically based on message load and network fairness.

% Requires \usepackage{booktabs, tabularx}
\begin{table}[t]
\centering
\footnotesize
\caption{Parameters for ebit generation}
\label{tab:assumptions}
\begin{tabularx}{\columnwidth}{@{}llX@{}}
\toprule
Parameter & Value & Basis \\
\midrule
Cavity, [g, $\kappa_{sc}$, $\kappa_{wg}$] & [1.0, 0.5, 2.35] $\times 2 \pi$ GHz & Designed proposal parameters \\
Mode matching, $\eta_\mathrm{mm}$ & 0.75 & 0.65 shown, $\sim$0.9 theory \cite{Menon2024AnImaging} \\
Fiber coupling, $\eta_\mathrm{fiber}$ & 0.75 & Spec \\
Herald cutoff, $\eta_\mathrm{herald}$ & 0.9 & \cite{Menon2020NanophotonicInterface, Huie2021MultiplexedCavities} \\
Per interface, $\eta$ & 0.4 & Product of the above \cite{ShankarGirijavallabhanMenon2025CombiningNetworks}\\
Herald misread & $10^{-5}$ & Spec \\
Atom conveyor speed & 0.3\,$\mu$m/$\mu$s & \cite{Menon2024AnImaging} \\
Transport speed & 1\,$\mu$m/$\mu$s & Spec \cite{Sinclair2025Fault-tolerantArrays} \\
Single-qubit gate error & $p_{2Q}/10$ & Assumption,1e-4 \cite{evered2023high} \\
Measurement time & 500\,$\mu$s & \cite{Bluvstein2024LogicalArrays} \\
Rearrangement time & 0.25\,ms & \cite{Bluvstein2022AArrays} \\
Herald time & 1\,$\mu$s & Calculated from cavity params \\
Measurement error & $10^{-4}$ & \cite{Manetsch2025} \\
Field of view & 1.5\,mm & \cite{Manetsch2025} \\
\bottomrule
\end{tabularx}
\end{table}

These estimates are close to the current experimental state of the art (Table~\ref{tab:assumptions}) and leave several clear avenues for improvement. At this operating point, the rate is set not by photon loss but by the node's $\sim$1.5\,ms batch cycle, limited by the three 500\,$\mu$s fluorescence readouts in the measurement zone, so faster readout schemes would directly raise the rate. Beyond this, the rate grows linearly with the number of multiplexed cavities and atoms and quadratically with $\eta$ (Figure~\ref{fig:exp_numerics}c), so a larger field of view, improved cavity and fiber coupling, and lower-loss hollow-core fiber all offer direct gains. The fidelity can be increased by reducing $p_{atom\text{-}photon}$, for example with polarization-protected level schemes \cite{Menon2020NanophotonicInterface} and phase-stabilized lasers that leave predominantly phase error, which purification removes efficiently (Figure~\ref{fig:exp_numerics}b), and by further improvements in two-qubit gate fidelity.

\begin{figure}[ht]
\includegraphics[width=0.95\columnwidth]{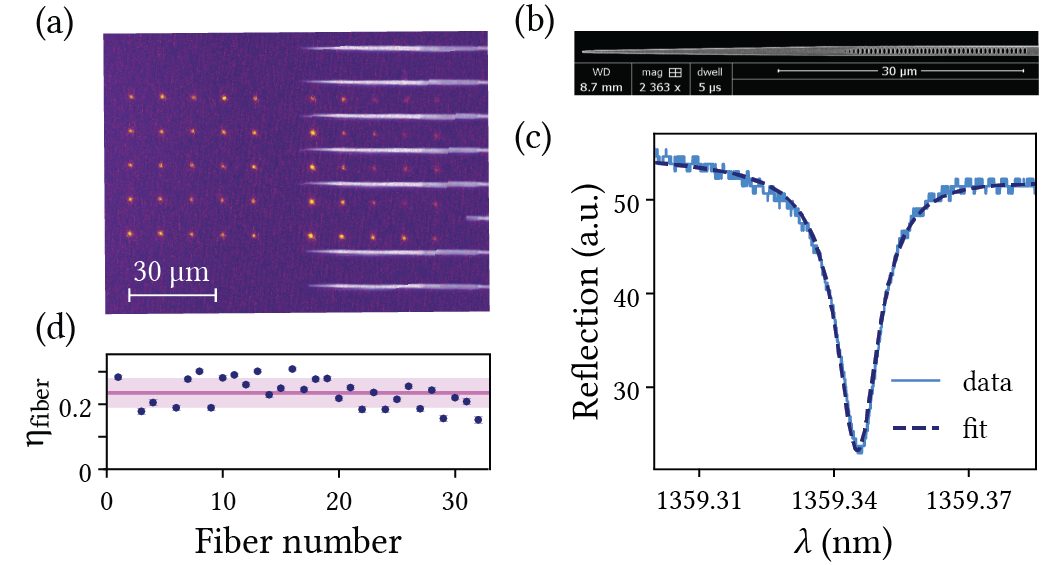} \caption{Experimental progress. (a) Atom array and cavity array. (b) SEM image of a nanophotonic device. (c) Cavity reflection spectrum with Lorentzian fit. (d) Fiber coupling efficiency for 30 devices coupled simultaneously.}\label{fig:exp_results}
\end{figure}

Although these rates and fidelities rely on projected capabilities of such a processor, we have made experimental progress toward the basic building blocks. Figure~\ref{fig:exp_results}a shows an atom array alongside an array of fabricated cavities with a pitch of 12\,$\mu$m, and Figure~\ref{fig:exp_results}b is an SEM image of a single device, showing the tapered waveguide and the hole-patterned cavity region. Our current chip hosts 150 devices, all of which could be addressed with a 2\,mm field-of-view objective, a modest extension from the state-of-the-art \cite{Manetsch2025}. The cavity resonance in Figure~\ref{fig:exp_results}c lies at $\lambda_0 = 1359.345(2)$\,nm, close to the atomic transition, with $\Delta\lambda_\mathrm{FWHM} = 11.42(3)$\,pm, corresponding to a quality factor $Q = \lambda_0/\Delta\lambda_\mathrm{FWHM} \approx 1.19(3)\times10^5$ and a linewidth $\kappa = 2\pi c\,\Delta\lambda_\mathrm{FWHM}/\lambda_0^2 = 2\pi\times1.85(4)$\,GHz which approaches the proposed cavity design parameters in Table~\ref{tab:assumptions}. In Figure~\ref{fig:exp_results}d, we show preliminary results of simultaneously coupling light from 30 devices into 30 fibers with an average fiber coupling efficiency of $0.23 \pm 0.05$, which could dynamically connect the \locoswitch{} node to local users and the neighboring repeater. We are now working toward observing high-cooperativity atom--cavity coupling in a single cavity, and toward multiplexed entanglement generation across many cavities and many atoms.

\subsection{Logical Error Simulations}\label{sec:eval-ler}

\begin{figure}[t]
\centering
\includegraphics[width=\linewidth]{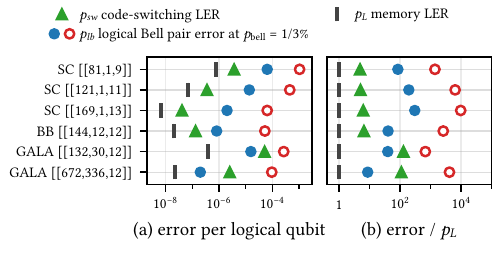}
\caption{Logical error rate under circuit-level simulations}
\label{fig:betas}
\end{figure}

The error incurred from the logical primitives is obtained using accurate circuit-level simulations, which we report in Figure~\ref{fig:betas}. We follow the noise model of Stack et al.~\cite{stack2026transversal}, where a depolarizing noise of strength $p$ is placed after every one- and two-qubit gate, including measurement and reset. We also apply a depolarizing noise of strength $p_{\rm bell}$ is applied to a pair of ebits. We sample the circuits with Stim~\cite{Gidney_2021} and decode them with Relay-BP~\cite{müller2025improvedbeliefpropagationsufficient}, falling back to BP-OSD~\cite{Panteleev_2021} on the shots where it does not converge. A logical Bell pair is prepared by the transversal non-local CNOT: each node runs four rounds of syndrome extraction on its block, every physical CNOT between the two blocks consumes one physical ebit, three more rounds follow, and both blocks are read out, so that a failure is any of the block's $k$ logical Bell pairs read wrong. Code-switching, like the logical Bell measurement that joins two logical ebits, is a joint $\bar Z\bar Z$ measurement between two blocks using a surgery adapter, which we simulate with the same noise and decoders and compare with the memory of the same blocks over the same rounds. Direct simulation at $p_{phys} = 10^{-3}$ is costly, as $O(m/p_L)$ shots are required in expectation to sample $m$ errors, which for some high-rate qLDPC codes can require billions of shots per data point. We instead simulate the code-switching logical error rate at $p_{phys} = 1.75\times 10^{-3}$ to $3\times 10^{-3}$ and report the extrapolated LER at $10^{-3}$ using the exponential form fit $p_L = p_{phys}^{d/2}\exp{(a + bp+cp^2)}$, as is done in \cite{bravyi2024high}.

\subsection{Network Simulations}\label{sec:eval-routing}
We evaluate \locoswitch{}\, on three families of networks: the simple chain of Figure~\ref{fig:logical} (toy$N$, $N$ users per side), the Abilene backbone of Figure~\ref{fig:hero} (abilene$N$: 11 cities, 14 intercity links of 14,082\,km~\cite{zhang2004abilene}, $N$ users per city) and a data-center tree of~\cite{zhang2025switchqnet} (spine10$\times N$: 10 racks of $N$ modules behind leaf switches under one spine). A switch's metro or in-rack links generate $10^{5}$ physical ebits per second, and each intercity or inter-rack link $10^{4}$; the links of a switch share its cavities, so the ebits they consume per second, each over its link's rate, sum to at most one. Bell infidelities are $p_{\rm bell}=1\%$ on every link (or 3\% where stated). Requests for a logical ebit of each class arrive as Poisson traffic at $R$ logical ebits per second, uniform over the user pairs, or as derived from two application traces from file verification \cite{supercheq} and distributed Shor's algorithm \cite{XueCovey2026}, and the baseline, the regenerating repeater baseline, and \locoswitch{} share the same modeling assumptions, available paths, and QEC protocols of Section~\ref{sec:framework}.

The batched plan reports the logical throughput capacity as the largest load that some window of 1\,ms to 10\,s sustains. The online protocol then runs between serving an offered load of 0.1 to 0.95 of that capacity at 120\,s of simulated time to test for stability, over 8 seeds on Abilene with 30 users per city and 3 for other topologies. We say a load is \emph{stable} when every seed delivers at least 97\% of the offered load and the mean latency is at most 1.5 times that at the lowest load, where a request's latency is its completion time minus the time the request is received. We report the latency as the mean latency over the seeds at loads up to half capacity. The simulator costs 13.3\,$\mu$s per event in the median over the Abilene runs, and it is released with this paper.

\subsubsection{The Load--Latency Frontier}\label{sec:eval-frontier}
In our batched plan on Abilene with 30 users per city, the baseline sustains 169 (41) logical ebits per second at FASQ target with 1\% $p_{\rm bell}$ (EFT with 3\% $p_{\rm bell}$), while \locoswitch{} sustains 4,362 (2,328) logical ebits per second, or 25.8$\times$ (56.2$\times$) better, respectively. The regenerating repeater sustains 169 at FASQ, where distilling the links buys nothing against the end-to-end check, and 86 at EFT with 3\% $p_{\rm bell}$, so \locoswitch{} gains 25.7$\times$ and 27.2$\times$ against it. Pooling and re-encoding account for the gains: in our event-driven simulation at half the baseline's capacity the baseline's levels are filled to 0.13 and it uses 120 ebits of the intercity links per delivered logical ebit, against full blocks and 5.0 for \locoswitch{}. At EFT with 3\% $p_{\rm bell}$ the QEC protocol differs as well, a GALA-672 block on 3\% pairs fails once in thirty, and end-to-end pairs over five links arrive with an error near 14\%, so the baseline need two levels of distillations, whereas \locoswitch{} distills raw ebits once per link, pooled, and once more at the route's edge switches, pooled over the city pair, at 8 ebits of the intercity links per delivered logical ebit against the baseline's 470, or about 60$\times$ fewer.

Online, without a solver, the baseline is stable up to 0.8 of its maximum capacity at FASQ, 136 logical ebits per second, and 0.7 at EFT with 3\% $p_{\rm bell}$, 29, and \locoswitch{} to 0.8 of its own at both classes, 3,488 and 1,862, delivering 97 to 99\% of the offered load up to that point. The regenerating repeater, over 3 seeds, is stable to 0.8 of its own capacity at both classes. The three designs therefore realize the same share of their plan capacities online. We also note that the latency of \locoswitch{} tends to be worse than baselines when the network is under saturated because its logical ebit passes through scarce links as high rate blocks, which need longer decoding time and code-switching operations. 

\subsubsection{Logical Throughput Gains}\label{sec:eval-topologies}

\begin{figure}[t]
\centering
\includegraphics[width=\columnwidth]{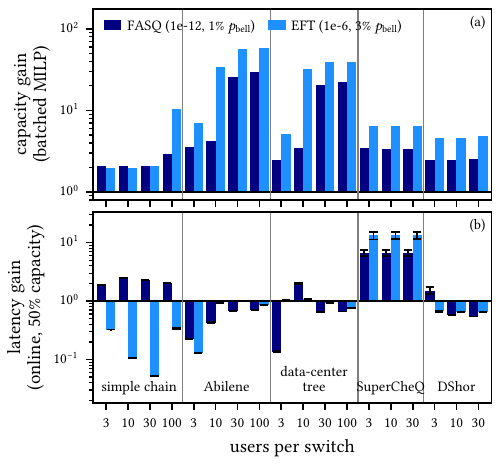}
\caption{Throughput and latency comparison over different network topologies and traces, higher is better. (a) The capacity gain of \locoswitch{} over the baseline, using the batched MILP algorithm. (b) The ratio of the two designs' mean online latencies at half the baseline's plan capacity, using the online algorithm; above 1, \locoswitch{} delivers sooner. }
\label{fig:topologies}
\end{figure}

\begin{figure}
    \centering
    \includegraphics[width=\linewidth]{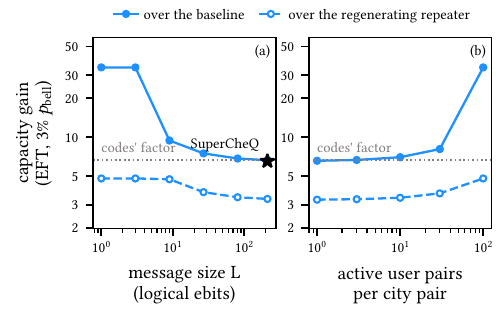}
    \caption{\locoswitch{} gain based on message size (a) and number of users per city (b).}
    \label{fig:gainvfill}
\end{figure}

The gain in capacity holds on every family (Figure~\ref{fig:topologies}a): in our batched plan it is 2.1$\times$ to 29.6$\times$ at FASQ with 1\% $p_{\rm bell}$ and 2.0$\times$ to 58.3$\times$ at EFT with 3\% $p_{\rm bell}$ over the baseline, and the same at FASQ and 1.7$\times$ to 29.2$\times$ at EFT over the regenerating repeater, which gains 1.18$\times$ to 2.06$\times$ over the baseline at EFT with 3\% $p_{\rm bell}$, 7.16$\times$ on Abilene with 10 users per city, where the baseline's twice-checked pairs sustain 68 logical ebits per second, and nothing at FASQ or at EFT with 1\%. We also report the gain taken from application traces of file verification via SuperCheQ\cite{supercheq}, achieving 3.37–3.43$\times$ gain at FASQ targets and 6.49–6.53$\times$ gain at EFT, and the communication trace of a modular computer implementing the distributed Shor's algorithm \cite{XueCovey2026}, achieving 2.49–2.54$\times$ and 4.63–4.79$\times$ at FASQ and EFT targets. 

\emph{The gain grows with the users per switch at both classes across topology and decreases with average message size}, as shown in Figure~\ref{fig:gainvfill}. Indeed, \locoswitch{} can reach the batch size required to fill high-rate blocks more easily since it can pool message requests of multiple users, even when each user individually cannot fill a high-rate block; so when there are not enough users, or when each user already have a large message to send that they can afford the large batch sizes themselves, the gain is less pronounced. On Abilene at EFT it grows from 6.9$\times$ with 3 users per city to 34.3$\times$ with 10 and 58.3$\times$ with 100, and at FASQ it stays at 3.5$\times$ to 4.2$\times$ with 3 to 10 users per city, where a user pair's requests still fill its nine-output levels within their storage limit, and rises to 25.6$\times$ and 29.6$\times$ with 30 and 100, where they do not; the data-center tree behaves alike, 2.5$\times$ to 22.4$\times$ at FASQ and 5.1$\times$ to 39.7$\times$ at EFT from 3 to 100 modules per rack, whereas the simple chain, where each user pair fills its own blocks, stays at 2.0$\times$ to 2.9$\times$.

Online, at half the baseline's plan capacity in our event-driven simulation (Figure~\ref{fig:topologies}b), the baseline delivers sooner on most settings, in 0.13 to 2.90\,s against 1.99 to 4.20\,s for \locoswitch{}, since at that load its distilled pairs go end to end into the user pair's levels while combined logical ebits of \locoswitch{} require code switching and longer decoding time on scarce link; \locoswitch{} delivers sooner, by 1.87$\times$ to 2.45$\times$, on the simple chain at FASQ and on the data-center tree with 10 modules per rack, where the baseline's per-user-pair pools wait for their levels, and the two are within 10\% of each other on Abilene with 10 and 30 users per city and on the tree with 3 to 30 modules per rack at EFT.

\subsubsection{Ablations}\label{sec:eval-mechanisms}

\begin{figure*}[t]
\centering
\includegraphics[width=\textwidth]{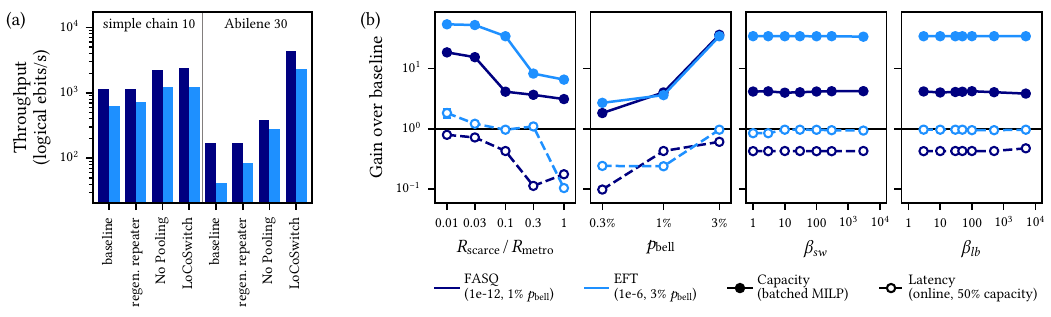}
\caption{(a)~Ablation studies: the throughput capacity of the batched plan for four designs, the baseline, the regenerating repeater, \locoswitch{} without pooling and \locoswitch{}, on the simple chain with 10 users per side (left) and on Abilene with 30 users per city (right) (b)~Sensitivity: the gain over the baseline on Abilene with 10 users per city, one knob at a time: the scarce links' rate relative to the metro links, $p_{\rm bell}$, $\beta_{sw}$ and $\beta_{lb}$. Both plots are over two target classes, FASQ with 1\% $p_{\rm bell}$ (dark) and EFT with 3\% $p_{\rm bell}$ (light), with logarithmic axes.}
\label{fig:ablation-sensitivity}
\end{figure*}

We remove the mechanisms of the gain one at a time in our batched plan, on the simple chain with 10 users per side and on Abilene with 30 users per city (Figure~\ref{fig:ablation-sensitivity}a). On the simple chain, where the logical load is shared amoung a few users and each user pair can fill a high rate block by themselves, pooling adds little, and the baseline and the regenerating repeater sustain 1,151 logical ebits per second at FASQ with 1\% $p_{\rm bell}$, \locoswitch{} without pooling 2,222 and \locoswitch{} 2,399, and at EFT with 3\% $p_{\rm bell}$ 619, 729, 1,221 and 1,227. On Abilene with 30 users per city, where a block launched for one request on a $10^{4}$ link costs its full 672 ebits, pooling across requests becomes the main mechanism. There, the four designs sustain 169, 169, 383 and 4,303 at FASQ and 41, 86, 278 and 2,328 at EFT, so that pooling multiplies the capacity by 11.2$\times$ and 8.4$\times$, whereas distilling every link, which the regenerating repeater adds, helps only at EFT, where the ebits are poor, by 2.1$\times$. At full fill the scarce links cost the baseline 35.2 physical ebits per logical ebit at FASQ against 9.98 for \locoswitch{}, 128 against 19.2 at EFT with 3\% $p_{\rm bell}$ and 32.3 against 9.93 at EFT with 1\% $p_{\rm bell}$ on Abilene with 10 users per city; this ratio of 3.5, 6.7 and 3.3 is the part of the gain that the codes give, and the rest is the fill of the baseline's per-user-pair levels.

\subsubsection{Sensitivity}\label{sec:eval-sensitivity}
We also test sensitivity to hardware and logical assumptions in Figure~\ref{fig:ablation-sensitivity}b, and show that the gain depends most heavily on the scarce link and on the error of the pairs. In our batched plan on Abilene with 10 users per city, raising the scarce links' rate from 0.01 to 1 of the metro links' lowers the gain from 54.1$\times$ to 6.5$\times$ at EFT with 3\% $p_{\rm bell}$ and from 18.5$\times$ to 3.1$\times$ at FASQ, and $p_{\rm bell}$ of 0.3\%, 1\% and 3\% gives 2.7$\times$, 3.6$\times$ and 34.3$\times$ at EFT and 1.8$\times$, 4.0$\times$ and 36.2$\times$ at FASQ, since at 3\% the baseline's end-to-end pairs need two checks. Over the noise biases $\beta_{sw} = p_{sw}/p_L$, $\beta_{lb} = p_{lb}/p_L$, the gain is relatively flat (the two right-hand panels of Figure~\ref{fig:ablation-sensitivity}b), because a final distillation at the edge switches and at the users can very efficiently remove the error of the switches' surgeries and of the blocks' preparation before it reaches the delivered logical ebit, and the plan is bound by the physical ebits that a logical ebit costs on the scarce links. 

\section{Discussion and Conclusion}\label{sec:conclusion}

\textit{\textbf{Limitations.}} 
To make the network simulations scalable to a large collection of protocols based on different code families and technologies in the literature, \emph{locosim} made several key assumptions to model physical and logical errors of the network and the processing nodes. The physical layer implementation couples a high-fidelity, scalable neutral atom array processor to a high-rate nanophotonic cavity-array network interface to facilitate user, switch, and repeater nodes. Under assumptions close to the current experimental state of the art, with hollow-core fiber and 99.99\% two-qubit gates at the repeater nodes, the \locoswitch{} node supplies ebits with $p_{bell} \leq 10^{-2}$ at $10^{4}$\,ebits/s across a 500\,km link, which is suitable for the high-rate codes. Secondly, we made conservative assumptions based on a sub-optimal BP-OSD decoder, which is likely several orders of magnitude worse than the SoTA decoders such as ML decoders \cite{gu2026scalableneuraldecoderspractical} and hierarchical decoders \cite{bhardwaj2026highrateqldpcprocessors,zhao2026towards}. Indeed, \cite{bhardwaj2026highrateqldpcprocessors} showed that the LER of surgery on high-rate qLDPC codes can reach $\sim 10^{-10}$ (or $\sim 16\times p_L$) with sub-millisecond decoding time, compared to our assumed $\sim100\times p_L$ for GALA and $59[ms]$ decoding time. Furthermore, if we assume that $p_{phys} = 2\times 10^{-4}$ is achievable, \cite{bonillaataides2025constant} have shown that up to $2\times 10^{-17}$ teleportation error rate is projected to be possible with qLDPC codes at $p_{bell} = 1\%$, whereas we assumed $\leq 10^{-6}$ with $p_{phys} = 10^{-3}$.  Finally, we note that while the error models assumed are reasonable for the atom system considered here, where idle errors are orders of magnitude below gate errors~\cite{bhardwaj2026highrateqldpcprocessors, zhao2026towards, yang2026designercodesgalacompact} and the syndrome extraction for surgery is fast with AOD-compatible moves~\cite{zhou2026parallelalgebraicsurgeryqldpc}; they do not extend to superconducting qubits, where the same circuits need more depth and connectivity~\cite{yoder2025tour}. It is therefore important to update the logical and physical assumptions with tailored modeling to assess the gains achievable from a message-combining switch, before we can generalize LoCoSwitch to other quantum computing systems such as superconducting and spin qubits.

\noindent\textit{\textbf{Future directions.}} While logical throughput, error rates, and latency are important targets for useful quantum networks, our modeling and simulator open up many exciting future directions, merging protocols and insights from classical networking architecture to a future quantum network. First is routing and congestion control beyond AIMD\cite{jain1984quantitative}. Up to a point, a larger batch of logical information requires fewer ebits per message. This creates a non-linear trade-off between efficiency and fair allocation of the bandwidth between users: a heavier load from a user does not necessarily reduce a link's capacity to serve others, since the larger batch also unlocks more efficient encoding. Secondly, resilience against switch or link failures is also important to investigate further: while high-rate codes can improve throughput and end-to-end latency, a lost high-rate code may be more expensive to recover. Finally, as secure quantum communication is an important application for a quantum network, it is also interesting to investigate how the capacities of error correction protocols--for example, the weight $d$-protection against bit/phase flip errors-- change under adversarial noise. We release \emph{locosim} with these questions in mind, and hope other features such as a failure model, a bandwidth allocation rule, or an adversarial analysis can be added alongside its routing and pooling protocols to facilitate future studies.

\noindent\textit{\textbf{Conclusion.}} In this work, we presented \locoswitch, a \locoswitchfull\, that can encode and re-encode ebit requests from its multiple users into more efficient ultra-high-rate quantum error-correcting codes based on traffic and link capabilities and perform QEC as logical ebits are routed through a quantum network, improving its logical throughput over baselines that optimize end-to-end error correction protocols and physical-level entanglement distribution/distillation separately. Such a switch faced three gaps across all layers of the network stack, and we addressed each: at the physical level, a neutral-atom node that stores and computes on logical qubits while its nanophotonic cavities generate entanglement at high rate; at the logical level, a list of codes and fault tolerant protocols for code-switching, and entanglement swapping, with logical error rates from circuit-level simulations; and at the network level, an ebit-efficient logical entanglement routing protocol that tries to choose the best routing path along with the QEC protocol at each link, solved optimally using a MILP algorithm over batched windows, and by a header-driven streaming algorithm requiring no centralized solvers. Against the baseline approach where each end-to-end users pick the best published protocol \cite{pattison2025constant, bonillaataides2025constant, stack2026transversal}, \locoswitch{} achieves 2.0$\times$ to 58.3$\times$ better logical throughput across 12 settings over three topologies and two application traces and up to 27.2$\times$ that of a regenerating repeater baseline on a network modeled after existing hardware specs, with better gains at more users per link and more bias between the rate of scarce and metro links. In our event-driven simulation of the same setting, the online protocol 0.8 of that capacity, consistent with the baseline, showing clear evidence of advantage. 

More broadly, we show that routing logical information in a quantum network is a logical-level problem that must consider the information encoding at each link for both efficiency and error correction capacity, while existing compartmentalized routing protocols consider only physical ebits or a homogeneous family of codes, which can congest scarce links and reduce overall capacity. Quantum networking, with point-to-point links and networks of a few nodes~\cite{hermans2022teleportation,pompili2022experimental,knaut2024entanglement,Stolk2024Metropolitan-scaleQubits}, is at roughly the stage that classical networking reached before the Internet's protocols were settled~\cite{baran1964distributed,cerf1974protocol}; its priorities and guiding principles are still taking shape. We give \locoswitch{} and \emph{locosim} with such questions in mind, and hope it will facilitate future research along this direction.
\\

\section{Acknowledgments}

WY acknowledges support by the NSF GRFP Program under Grant No. 2023360346.  A.R.K.\ acknowledges the support from the European Commission for a Marie Sk\l{}odowska-Curie individual fellowship No.~101269962 (MechNanoCav\_for\_QNN). J.C.B.\ and D.G.\ acknowledge support from the Austrian Science Fund (FWF) through the Special Research Programme Qnnect (SFB F1012). H.B.\ acknowledges support from the NSF CAREER program (NSF Award 2238860). FTC is the Chief Scientist for Quantum Software at Infleqtion. This work is funded in part by the STAQ project under award NSF Phy-232580; in part by the US Department of Energy Office of Advanced Scientific Computing Research, Accelerated Research for Quantum Computing Program; and in part by the NSF Quantum Leap Challenge Institute for Hybrid Quantum Architectures and Networks (NSF Award 2016136), in part by the NSF National Virtual Quantum Laboratory program, in part based upon work supported by the U.S. Department of Energy, Office of Science, National Quantum Information Science Research Centers, and in part by the Army Research Office under Grant Number W911NF-23-1-0077. The views and conclusions contained in this document are those of the authors and should not be interpreted as representing the official policies, either expressed or implied, of the U.S. Government. The U.S. Government is authorized to reproduce and distribute reprints for Government purposes notwithstanding any copyright notation herein.

\noindent\textit{\textbf{AI Statement.}} We used Claude Fable 5.1 and Opus 5.5 to implement \emph{locosim} based on preliminary code written by the authors, generate the map in Figure 1 and Figure 4a, as well as edit the figure formatting. We used Grammarly and Opus 5.5 for copyediting. Authors have reviewed and take full responsibility for all results.

\bibliographystyle{IEEEtran}
\bibliography{refs}

\end{document}